\documentstyle[12pt]{article}
\newcommand{\be}{\begin{equation}}
\newcommand{\ee}{\end{equation}}
\newcommand{\ba}{\begin{eqnarray}}
\newcommand{\ea}{\end{eqnarray}}
\newcommand{\baa}{\begin{eqnarray*}}
\newcommand{\eaa}{\end{eqnarray*}}
\newcommand{\bb}{\begin{thebibliography}} \newcommand{\eb}
{\end{thebibliography}}
\newcommand{\ci}[1]{\cite{#1}}
\newcommand{\bi}[1]{\bibitem{#1}}
\font\twelve=cmbx10 at 15pt
\font\ten=cmbx10 at 12pt

\begin{document}

\begin{titlepage}

\begin{center}

{\ten Centre de Physique Th\'eorique - CNRS - Luminy, Case 907}

{\ten F-13288 Marseille Cedex 9 - France }

{\ten Unit\'e Propre de Recherche 7061}

\vspace{1 cm}

{\twelve ON THE $G_2$ MANIFESTATION
FOR
LONGITUDINALLY  POLARIZED PARTICLES}\footnote{ This
 research was partly supported by INTAS (International Association
for the Promotion of Cooperation with Scientists from the
Independent States of the Former Soviet Union) under Contract nb 93-1180.}

\vspace{0.3 cm}

{\bf J.
SOFFER and O.V.
TERYAEV}\footnote{Permanent address : Bogoliubov Laboratory of Theoretical
Physics,  Joint Institute for
 Nuclear Research, Dubna, Head Post Office, P.O. Box 79,  101000
Moscow, Russia}

\vspace{1.5 cm}

{\bf Abstract}

\end{center}

The contribution of the $G_2$ structure function
to  polarized  deep  inelastic scattering is slightly redefined
in order to avoid   kinematical zeros. Its strong
$Q^2$-dependence implied by the Burkhardt-Cottingham (BC) sum
rule naturally explains the sign change of the generalized
Gerasimov-Drell-Hearn (GDH) sum rule.  The status of the BC sum
rule and implications for  other spin processes are discussed.

\vspace{2 cm}


\bigskip

\noindent February 1994

\noindent CPT-94/P.3014
\bigskip

\noindent anonymous ftp or gopher: cpt.univ-mrs.fr

\end{titlepage}

\baselineskip 24pt
\section{Introduction}

The structure functions $G_1$ and $G_2$ describing the spin-dependent
 part
of deep-inelastic scattering were discussed by Feynman in his
celebrated lectures \cite{Fey} and the
definition he used is very appealing due to the simple partonic
interpretation of the
dimensionless function $g_1(x)$, the only one which survives in the
scaling limit for the longitudinal
polarization case. However, when one studies the relatively low-$Q^2$
region, the alternative definition of structure functions, proposed by
Schwinger a few years later
\cite{Sch} appears to be more useful. In
fact, it was recently applied to explain the strong
$Q^2$-dependence of the generalized
Gerasimov-Drell-Hearn (GDH) sum rule\ci{SoTe}. The crossing point
where this integral turns to zero
is, to a large extent, determined by the BC sum rule. This approach
was criticized in the recent
paper \cite{Ji}, devoted to the interpolation between high and
low $Q^2$ and a comment \cite{WoLi} on ref.\ci{SoTe}
has put in focus  interesting problems related to the resonance
contributions.

In this article we present a systematic analysis of the  BC sum rule
and the $G_2$ manifestation in scattering with longitudinally polarized
particles. The basic definitions are introduced  in Section 2 and we
compare them. The GDH problem is analyzed in Section 3 and the
implications to the BC sum rule validity are presented in Section 4.
Section 5 is devoted to the elastic and resonance contributions and to
the discussion of the papers
\cite{Ji,WoLi}. In  Section 6 the possible manifestations of these
effects in  other spin-dependent processes are discussed, while our
conclusions are presented in  Section 7.

\section{Definitions of the spin-dependent structure functions}

To define the spin-dependent structure functions one should express
the antisymmetric part of the hadronic tensor $W^{\mu \nu}$ as a
linear combination of all possible Lorentz-covariant tensors. These
tensors should be orthogonal to the virtual photon momentum $q$, as
required by the gauge invariance, and they are linear in the nucleon
covariant polarization $s$ from a general property
of the density matrix. If the nucleon has momentum $p$, we have, as
usual,
$sp=0$ and $s^2=-1$. There are only two such tensors; the first one
arises already in the Born diagram
\be T_1^{\mu
\nu}= \epsilon^{\mu \nu \alpha \beta}
s_\alpha q_\beta
\ee
and the second
tensor is just
\be
T_2^{\mu \nu}=(sq) \epsilon^{\mu \nu \alpha\beta}p_{\alpha}q_{\beta}.
\ee
The scalar coefficients of these tensors are specified in a
well-known way

\ba
W^{\mu \nu}_A={-i\epsilon^
{\mu \nu  \alpha\beta}\over {pq}}q_{\beta}
(g_1 (x, Q^2)s_{\alpha}+g_2 (x,Q^2)(s_{\alpha}-p_{\alpha}{sq \over
{pq}}))=\nonumber \\
{-i\epsilon^ {\mu \nu  \alpha\beta}\over {pq}}q_{\beta}((g_1 (x,
Q^2)+g_2 (x,Q^2))s_{\alpha}-g_2
(x,Q^2)p_{\alpha}{sq \over {pq}}). \ea

Here we used the dimensionless functions $g_1$ and $g_2$ whose
$Q^2$-dependence,  in the scaling
region,  reduces to  rather weak scaling violations. However,
they can also be used for  low $Q^2$ : the $Q^2$-dependence is then
strong and non-calculable in perturbative QCD. One can easily recover
the conventional $G_{1}=g_1 M^2/pq$ and $G_2=g_2 M^4/(pq)^2$, but
$g_{1,2}$ appear to be  most convenient for our purposes.

In fact, considering the case of  {\it longitudinal polarization}, one
can easily transform (3) as follows:
$$
W^{\mu \nu}_A={i\over M}\epsilon^{\mu \nu}_{\perp}
(g_1(x,Q^2)-g_2(x,Q^2){Q^2 M^2
\over {(pq)^2}}),\eqno (3')
$$
where
$\epsilon_{\perp}$ is a two--dimensional antisymmetric tensor in
the hyperplane orthogonal to $p$ and $q$.
The $g_2$ contribution drops either if $Q^2=0$, or in the
scaling limit.
In the later case the suppression is not exact, and leads to a
higher-twist term $4g_2x^2 M^2/Q^2$. Note that the coefficient of
$g_2$ is of order unity in the resonance region and therefore the
manifestation of  $g_2$ for the elastic contribution \ci{Ji} is by
no means surprising.

The cancellation of $g_2$
is a direct consequence of the  definition assumed in (3). It allows
to describe the longitudinal polarization, which is kinematically
dominant (the common factor $(pq)^{-1}$ in (3) is absent in (3$'$)) ,
by the single structure function $g_1$. However, it confirms the fact
that both tensors (1) and (2) contribute and this fact
becomes important, if one is interested in the dynamical properties of
$g_1$ and $g_2$.

As it was mentioned above, the $T_1$ tensor is "simpler": it emerges
 in the Born diagrams and it is the natural candidate for the
application of the QCD sum rules method. Concerning the coefficient of
$T_2$, it is strongly restricted by the Burkhardt-Cottingham sum rule
and should be equal to zero "in average".

The most natural way to account for this difference is to take the
 coefficients of $T_1$ and $T_2$ (i.e. $g_1+g_2$ and $-g_2$, see
second line in eq.(3)) as {\it independent} \ci{Sch}. This definition
removes the mentioned kinematical zero of $g_2$ at $Q^2=0$. It seemed
to us
\ci{SoTe} that it is possible to use the same notations, just
remembering that it is impossible now to cancel the $g_2$ terms of
$T_1$ and $T_2$. However this may be  misleading and it seems better to
define
\be
g_{1+2}=g_T \equiv g_1+g_2.
\ee
The subscript $T$ is the reminder of the well-known fact that only
$T_1$ contributes to the transverse polarization case.

There is an important {\it physical} difference between the real and
virtual
 photon cases. In the later case it is possible to extract from the
experimental data {\it two} independent scalar functions (whatever
they are defined): one has just to measure the asymmetries for
longitudinally and transversely polarized nucleon. For real photon the
transverse asymmetry is equal to zero. However, there is no reason to
identify the longitudinal asymmetry with the contribution of $T_1$,
except  the kinematical zero in the standard definition. Moreover, it
may be of some help
to study the contributions of $T_1$ and $T_2$  separately.

\section{The GDH and BC sum rules}

The main problem with the generalized Gerasimov-Drell-Hearn
\ci{Ger,DH}
 sum rule is the following.
Consider the $Q^2$-dependent integral
\be
I_1(Q^2)={2 M^2\over {Q^2}} \int^1_0 g_1(x) dx \ee
which is defined for {\it all} $Q^2$. Note that the elastic
contribution  at $x=1$ is not included. One then recovers  at $Q^2=0$
the GDH sum rule
 \be I_1(0)=-{\mu_A^2 \over 4},
\ee
where $\mu_A$ is the nucleon anomalous magnetic moment in nuclear
magnetons. While $I_1(0)$ is always negative, its magnitude and sign at
large
$Q^2$ are determined by the $Q^2$ independent integral $\int^1_0 g_1(x)
dx$. For the proton it is positive, so one should expect for
$I_1(Q^2)$ a strong
$Q^2$-dependence and one can ask : what is its origin?

It is possible to decompose $I_1$ according to the contributions of
 the tensors
$T_1$ and $T_2$
\be
I_1=I_{1+2}-I_2,
\ee
where
\be
I_{1+2}(Q^2)={2 M^2\over {Q^2}} \int^1_0 g_{1+2}(x) dx,\qquad I_2(Q^2)=
{2 M^2\over {Q^2}} \int^1_0 g_2(x) dx. \ee

There are solid theoretical arguments to expect a strong
$Q^2$-dependence of $I_2$. It is the well-known Burkhardt-Cottingham
sum rule \ci{BC}, derived independently by Schwinger \ci{Sch} using a
rather different method, i.e.,
\be I_2(Q^2)={1\over 4}\mu G_M (Q^2)[\mu G_M (Q^2) - eG_E
(Q^2)], \ee
 where $\mu$ is the nuclear magnetic moment, and the $G$'s denote
the familiar Sachs form factors which are dimensionless and
normalised to unity at $Q^2=0$. For large $Q^2$ one can neglect the
r.h.s. and one  gets
\be
\int^1_0 g_2(x) dx=0.
\ee

This later equation  is often called the BC sum rule.
However, the elastic contribution was also present in their pioneer
paper and it is this contribution which generates the strong
$Q^2$-dependence of $I_2$. In particular,
\be
I_2(0)={\mu_A^2+e\mu_A  \over 4},
\ee
$e$ being the nucleon charge in elementary units, so in order to
reproduce  the GDH value, one should have

\be
I_{1+2}(0)={e\mu_A \over 4}\ .
\ee
Note that $I_{1+2}$ does not differ from $I_1$ for large $Q^2$
due to the BC sum rule, but it is {\it positive} in the proton case
and it is possible to obtain a  smooth interpolation between large
$Q^2$ and
$Q^2=0$ \ci{SoTe}.

This smooth interpolation seems to be very reasonable in the
framework of the QCD sum rules method. Then one should choose  some
"dominant" tensor structure to study the $Q^2$-dependence of its scalar
coefficient and $T_1$ appears to be a good candidate. This seems
also promising  from another point of view. It is not trivial to
obtain within the QCD sum rules approach the  GDH value at $Q^2=0$.
Since the r.h.s. of (12) is linear in $\mu_A$, it may be possible to
obtain it using the Ward identities, just like the normalization
condition for the pion form factor \ci{Rad}.

Concerning the neutron case, $I_{1+2}(0)=0$ and eq.(10) naturally
explain the small absolute value of $g_1^n$ compared to
$g_1^p$, as required by the Bjorken sum rule and confirmed by the
recent SLAC and SMC measurements, despite  the controversies between
these two collaborations \ci{SMC,SLAC}. The best agreement between
$Q^2=0$ and high
$Q^2$ is provided by the $SU(6)$ value $g_1^n=0$.

In the proton case to give a quantitative prediction for $I_1(Q^2)$
one needs some  parametrization
to interpolate  $I_{1+2}$ between $Q^2=0$ and high $Q^2$. The
simplest one \ci{SoTe} is
\be
I_{1+2}(Q^2)=\theta(Q^2_0-Q^2)({\mu_A \over 4}- {2 M^2Q^2\over
{(Q^2_0)^2}} \int^1_0 g_1(x) dx)+\theta(Q^2-Q^2_0) {2 M^2\over {Q^2}}
\int^1_0 g_1(x) dx.
\ee
The continuity of the function and of its derivative is guaranteed
with the choice $Q^2_0=(16M^2/\mu_A) \int^1_0 g_1(x) dx \sim 1GeV^2$,
where the integral is given by the EMC data. It is quite a reasonable
value to separate the perturbative and non-perturbative regions. As a
result one obtains a crossing point at $Q^2 \sim 0.2 GeV^2$, below the
resonance region \ci{SoTe}. It is interesting, that the slope of
$I_1^p$ at the origin agrees with the result obtained recently in the
framework of chiral perturbation theory \ci{Meis}.

This result is not sensitive to  $I_{1+2}$  both at low and high
$Q^2$, provided the behavior is smooth, but is very sensitive to the
validity of BC sum rule whose possible violations have been extensively
discussed in the literature  \ci{IoLi,Jaf}.

{}From another point of view, the saturation of the GDH sum rule by
the contributions of low-lying resonances leads to a crossing
point around $0.6-0.8 GeV^2$ \ci{BuLi}. A simultaneous assumption about
the validity of the BC sum rule yields dramatic oscillations of
$I_{1+2}$ \ci{WoLi}.
These two subtle points are the subject of the next two sections.

\section{New implications for the validity of the BC sum rule}

The first implication for the BC sum rule comes just from the
Schwinger derivation \ci{Sch}. It was obtained by using the
antisymmetry property in $\nu$, the virtual photon energy, of the
relevant invariant amplitude, for which one writes a double spectral
form. The antisymmetry property implies that the integration over all
the kinematical region, including the elastic contribution, should
give zero. This derivation is similar to the derivation of the scaling
form of the BC sum rule, in the framework of the QCD twist-3 approach
\ci{ET84}. In this case the BC sum rule arises also as a result of
the integration of an antisymmetric function over a symmetric region
\be
\int^1_0 g_2(x) dx={1\over {2\pi}}\int_{|x,y,x-y|\leq 1}dxdy\
{b^A(x,y)\over {x-y}}. \ee
Here $b^A(x,y)$ is the dimensionless quark-gluon correlator
 \ci{ET85}, proportional to the double spectral density in the scaling
region. Its symmetry in $x,y$ follows from $T$-invariance, just like
$\nu$ antisymmetry in Schwinger derivation. Note, however, that the BC
sum rule is not spoiled by violations of the $T$-invariance :
$b^A(x,y)$ should be replaced by the symmetrized combination
$(b^A(x,y)+b^A(y,x))/2$ (see below, Section 6).

One of the  possibilities for BC sum rule violation is
the {\it long-range} singularity $\delta(x)$ \ci{Jaf}.If $g_2$ contains
such a term proportional to
$\delta(x)$, which can never be observed experimentally, it will give a
non-zero contribution to the integral (10) and therefore will
violate the BC sum rule. Note that such a situation was
first noticed by Ahmed and Ross in their pioneer paper on  spin
effects in QCD
\ci{Ah}, where they also mentioned a similarity with the Schwinger
 term sum rule for the longitudinal structure function.
However, to obtain such a behavior, one also needs to have a
singularity  in $b^A$  e.g.
\be
b^A(x,y)=\delta(x) \phi(x,y)+\delta(y) \phi(y,x), \ee
where $\phi$ is regular function. Such a singularity should result
in  meaningless infinite single asymmetries, which may be generated
by the correlators \ci{ET85,StQu}.

Another implication for the BC sum rule comes from  checking  the sum
 rules (9),(12) in QED \ci{Mil}, as it was performed immediately after
Schwinger paper.  The result in QCD is the same apart from a trivial
color factor.

Note that in the leading approximation $\mu G_M$ is just $e$, while
 $\mu G_M-eG_E$ is the anomalous magnetic form factor. For high
$Q^2$ it provides the elastic
contribution to the BC sum rule which is decreasing like
$Q^{-2}logQ^2$. However it should decrease {\it faster} than $Q^{-2}$
to get the standard zero BC sum rule. Therefore, in massive on-shell
one-loop QCD the BC sum rule is violated like
$\alpha_s log Q^2$. Taking the leading approximation for $\alpha_s$,
one observes the cancellation of $log's$ and one obtains the answer
\be
\int^1_0 g_2(x) dx=-{C_F\over{2\beta_1}}, \ee
$\beta_1$ being the one-loop beta-function. The cancellation of
$log's$
 is very similar to the one providing the anomalous gluon contribution
to polarized DIS \ci{EST}. However,  higher-order corrections and
non-perturbative confinement effects should make the elastic
contribution rapidly decreasing.  In the standard application of
QCD factorization, both elastic and inelastic contributions are
included to all orders of perturbation theory. This allows to cancel
the most infrared singularities and in particular the $logQ^2$
mentioned above. The result of ref.\ci{Mil} means that the BC sum rule
is then valid due to the cancellation of elastic and inelastic
contributions. It contradicts the recent result of Mertig and van
Neerven
\ci{Mer-vN} who found, that the partonic BC sum rule is violated for
both the
$\overline{MS}$ and the on-shell renormalization schemes. Note that
the later coincides with the calculations discussed above. The only
difference is that the quark mass $m$ is just a regulator of collinear
singularities, and only the terms contributing to the limit
$m\rightarrow 0$ were taken into account. However, this leads to  an
extra divergence when $x\rightarrow 1$, regularized by an extra
parameter
$\delta$, separating hard and soft gluons. As a result, a term
proportional to
$log\delta$ appears in the elastic contribution. However, the
generalized GDH sum rule tells us that it is proportional to the
anomalous magnetic moment, which is infrared stable. One may
conclude, that this BC sum rule violation is an artifact of the
approximation.

There is another possibility for the BC sum rule violation, namely
the non-scaling one \ci{IoLi}.
It comes from the different Regge asymptotics for the forward
helicity amplitudes, related to the structure functions $G_{1,2}$. The
Regge cuts from Pomeron, $P'$ and $A_2$ poles could spoil the
superconvergent sum rule for $G_2$ but this possibility
is not clear at all\ci{Pom}.

The presence of the non-scaling BC violation does not contradict
the main qualitative feature of our approach, namely, the strong
$Q^2$-dependence of the $g_2$ first moment (and therefore $I_2$
contribution to longitudinal polarization). Although the cut
contribution to the r.h.s. of
BC sum rule
is unknown, the
same contribution should be present in $I_{1+2}$, because the
Regge cuts do not contribute to $g_1$. Since it cancels out in $I_1$,
it is still possible to say, that the qualitative origin of the rapid
 variation of $I_1$ is the elastic contribution to $g_2$.

The later statement leads to a quantitative prediction namely,
the non-scaling violation of the BC sum rule should not affect the
position of the crossing point ($0.2 GeV^2$). Actually, it is natural
to decompose $g_{2,1+2}$ as follows

\be
g_2=g_2^{scale}+g_2^{el}+g_2^{cut},\quad g_{1+2}=g_{1+2}^{smooth}
+g_{1+2}^{cut}. \ee

The notations are obvious and since the first moment of
$g_2^{scale}$  is zero, the crossing point is determined by the
equation
 \be I_2^{el}(Q_{cross}^2)+I_2^{cut}(Q_{cross}^2)=
I_{1+2}^{smooth} (Q_{cross}^2)+I_{1+2}^{cut}(Q_{cross}^2). \ee

As it was mentioned above, the cut contributions to $I_2$ and
$I_{1+2}$ are equal. Moreover, to reproduce the GDH sum rule,
unaffected by the cuts, $I_{1+2}^{smooth}$ should approach the same
limit  $e\mu_A/4$ at $Q^2=0$. One can therefore repeat the arguments
of the previous section, relating the crossing point position to the
elastic contribution to the BC sum rule.

The goal  of future experiments is to check the BC sum rule
in the low-$Q^2$ region and the simultaneous study of  transverse
and longitudinal polarizations is  most appropriate for this
purpose. If the cut contribution is absent,  $I_{1+2}$ directly
measured for the transverse polarized target case, should change
smoothly, turning to the Schwinger value $e\mu_A/4$ at $Q^2=0$. The cut
contribution makes both $I_{1+2}$ and $I_2$  change rapidly in the
resonance region.

\section{The resonance and elastic contributions to the GDH and BC sum
rules}

The only way of experimental check of the GDH sum rule yet is its
saturation by the contributions of low-lying resonances. The central
role here plays the $\Delta(1232)$: it provides a significant amount
of GDH integral at $Q^2=0$ and gives a clear qualitative explanation
of rapid $Q^2$-dependence\ci{IoLi}. The $\Delta$ photoproduction is
dominated by the magnetic dipole form factor, leading to a negative
$I_1$. The sign change is just related  to the fast decrease of the
$\Delta$ contribution.

In order to compare this picture with our approach we have separated
the $\Delta$ contribution to $I_{1+2}$ and $I_2$. To do this we
just calculated the photoproduction Born diagram using the well-known
expressions for the covariant form factors $G_M$, $G_E$ and $G_C$. The
resulting expression, obtained with the help of FORM program \ci{JVer}
for symbolic computations, is rather lengthy but it has a remarkable
property: the leading $G_M^2$ term contains {\it only} the tensor
$T_2$ at any $Q^2$. This fact is confirmed if one performs the
contraction with the virtual photon density matrix: in particular, if
one takes the standard definition with the kinematical zero at
$Q^2=0$, the result should be attributed to $g_1$. In our approach the
nonzero
$g_1$ is due to the absence of $T_1$ since $g_1=-g_2$.

This result supports qualitatively our main conclusion, namely,
that the strong $Q^2$-dependence of GDH sum rule should be attributed
to
$g_2$. It also naturally explains the saturation  by $\Delta$ of the
neutron GDH sum rule in the $SU(6)$ limit ($I_{1+2}(0)=0$) as seen in
Section 3. The quantitative difference between our result and that of
the "resonance" approach may be explained in  different ways. Suppose,
we perform such a decomposition for the whole resonance contribution.
It is a more consistent approach than to treat simultaneously the
resonance approximation for $I_1$ and the fundamental BC sum rule for
$I_2$. Since the covariant form factors for higher resonances are
unknown, the best way may be to calculate the transverse asymmetry,
using the helicity amplitudes (although one needs to know more
 about it than for the longitudinal one). If, nevertheless,
such a decomposition is performed, three possibilities arise:

i) The results for $I_2$ coincide, but for $I_{1+2}$ they are
different,  possibility
suggested in \ci{WoLi}. If one really obtains this result in the
above mentioned manner, it would be the first check of the BC sum
rule.

ii) The results for $I_{1+2}$ coincide but they differ for $I_2$; in
this  case it would be either an argument in favour of the violation of
the BC sum rule or an indication for an additional contribution.
This situation is supported by the fact, that the kink structure
of $I_1$ at low $Q^2$ arises from the $\Delta$ contribution and
should be therefore present in $I_2$ and  the elastic
BC contribution is monotonic.

iii) Both $I_{1+2}$ and $I_2$ are different; this result would be
the most unclear and perhaps the most interesting.

When discussing the resonance contributions, it seems reasonable to
mention the elastic one and the main idea of ref.\ci{Ji} is
to interpolate between high and low $Q^2$ including the elastic
contribution. It is quite a different problem, because one can never
reach  $Q^2=0$ and a negative GDH value: the elastic contribution is
then absent for a trivial kinematical reason. Concerning the objections
of this paper, all of them are based on the standard definition of
$g_2$ with the kinematical zero.

\section{Generalization to  other spin-dependent processes}

The main point of our analysis is the decomposition of the
longitudinal polarization contribution into two pieces, one of which
is directly related to  $g_2$. It seems very easy to perform such
a decomposition for an arbitrary hard process with longitudinally
polarized particles.

One may also apply the QCD twist-3 approach \ci{ET84}, already
mentioned  in the previous section.
The spin-dependent part of any hard cross-section may be expressed
in a factorized form as
$$
d\sigma_s=\int dx tr[E(x)T(x)]+
\int dx_1dx_2 tr [E_{\mu}(x_1,x_2)T_{\mu}(x_1,x_2)],$$
with
\be T(x)=M(\hat
s \gamma^5 c_T(x)),
\ \ T_{\mu}(x_1,x_2)={M\over 2\pi}(\hat p\gamma^5s_{\mu}b_A(x_1,x_2)+
 i\gamma_{\rho}\epsilon^{\rho \mu
\alpha\beta}s_{\alpha}p_{\beta}b_V(x_1,x_2))\ .
\ee
$E^\mu(x_1, x_2)$ are density matrices of on-shell quark and quark
gluon perturbative QCD diagrams.
Here $c_T(x)$ and $b_{A,V}(x_1, x_2)$ are "ordinary" transverse
polarized quark distributions and quark-gluon correlators,
respectively. The two-argument distributions
$b_A$ and $b_V$
are real, dimensionless and they possess symmetry properties which
follow from $T$-invariance, i.e.

\begin{eqnarray}
b_A(x_1,x_2)=b_A(x_2,x_1),\qquad b_V(x_1,x_2)=-b_V(x_2,x_1).
 \end{eqnarray}

Note that only transverse distributions occur in  the natural
twist-3 basis. In the longitudinal polarization case $s^\mu=p^\mu/M$,
from eq.(19) it is clear that only the correlator $b_A$ contributes.
If one makes use of the sum rule
derived in \ci{ET84} in terms of the longitudinal quark distribution
$c_L(x)$

\be
\int dx (c_L(x)-c_T(x))\sigma(x)=
{1\over{2\pi}}\int dx_1dx_2 b_A(x_1,x_2) {{\sigma(x_1)-\sigma(x_2)}
 \over {x_1-x_2}}, \ee
where $\sigma(x)$ is an arbitrary test function, one  obtains the
simple expression

\be
d\sigma _{s,L}=\int dx tr[\hat p \gamma ^5E(x)c_L(x)]\ . \ee
The sum rule (20), whose natural consequence when $\sigma(x)\equiv 1$
is the BC sum rule, provides the decomposition of the longitudinal
spin-dependent quark distribution

\be
c_L(x)=c_T(x)+
{1\over{2\pi}}\int dy {b_A(x,y) \over {x-y}}\ . \ee
While the first piece ($c_T(x)$) is proportional to $g_1+g_2$ and we
have,
\be
g_1(x_B)+g_2(x_B)=c_T(x_B)+c_T(-x_B),
\ee
the
second one is related to $g_2$ (see eq.(14)).

Although these expressions are derived for hard processes,
one may also consider them as a {\it definition} of the parton
distributions in the soft region. As the flavour summation with
the target-depending weights is assumed, it seems interesting to study
the GDH problem for each flavour separately.

In the absence of accurate experimental information for  polarized
deep  inelastic scattering, one may use the indirect one,
provided by the Bjorken sum rule for the difference of the proton and
neutron structure functions \ci{IoLi}. For high $Q^2$, in our
notations,  it takes the form

\be
I_1^{u-d}(Q^2)={1\over 3}I_1^{p-n}(Q^2)={M^2\over {3 Q^2}} g_A. \ee
Here $g_A \sim 1.25$ is the axial $\beta$-decay coupling and
the limit at $Q^2=0$  is just

\be
I_1^{p-n}(0)={{\mu_n^2-\mu_p^2}\over 4}\ .
\ee

The important qualitative feature of these equations is the
fact that $I_1^{p-n}$ has the same (positive) sign at high $Q^2$ and
$Q^2=0$, so it allows to interpolate smoothly between the two regions
\ci{IoLi}. The transition value $Q^2_0$ (13) is, however, an order of
magnitude larger in this case. It is interesting to study the second
nonsinglet
$SU(3)$ combination \ci{OT92}. One should just change $u \rightarrow
d, d \rightarrow s, p \rightarrow \Xi _0, n \rightarrow \Xi _-$. Both
Bjorken and GDH sum rules change sign preserving the possibility for a
smooth interpolation. The transition $Q^2$ is lower in this case: it
is of the same order $1 GeV/c^2$, as for $I_{1+2}$ in the proton case.

The main qualitative consequence of the smooth behaviour of
non-singlet combinations is the following: the sharp $Q^2$-dependence
is likely to be attributed to the $SU(3)$-singlet channel. One may
expect that $g_2$, which is responsible for the strong
$Q^2$-dependence, also occurs mainly in the singlet channel.
This is the channel, where one has the EMC Spin Crisis and the gluon
anomaly
\ci{EST}. We performed the decomposition into $g_{1+2}$ and $g_2$
for the box diagram, giving rise to the anomalous gluon contribution.
It appears that the first moment, related to the axial anomaly, comes
only from $g_{1+2}$, while the contribution to $g_2$ is zero : the BC
sum rule is also respected for gluons. For higher moments the
situation is less trivial : the $g_{1+2}$ term does not contain the
logarithmic corrections, and  therefore may be related to the anomaly.
Its $x$-dependence is very simple
\be
E^g_{1+2}={\alpha_s\over\pi}(x-1)\ .
\ee
The {\it perturbative} $Q^2$-dependence of $g_1$ comes again from
$g_2$! However these relations between the anomaly, $g_2$, $\Delta$,
etc. require further investigations.

The sharp dependence of the singlet combination of quark densities
results in such a behaviour for each single density. One may ask, if
it can  be found in  processes different from  deep inelastic
scattering. The answer, in principle, is negative. Although formula
(23) is still valid, the new power corrections could appear which are
absent in the DIS case. If, however, these corrections
are small in comparison with the elastic BC contribution, one should
expect, e.g., a decreasing of the Drell-Yan longitudinal asymmetry at
$Q^2 \sim 0.2 GeV^2$.

\section {Conclusions}

The decomposition of the longitudinally polarized particle density
matrix into  two pieces seems to be very simple and natural. If one
starts from the very beginning with longitudinal polarization, the
particle is described by the single vector $p^\mu$, leading to the
single form factor $g_1\epsilon^{\mu \nu \alpha\beta}p_{\alpha}
q_{\beta}$ for deep inelastic scattering and to the single parton
density
$c_L$ for an arbitrary hard process. Starting instead with the
general case, described by the two vectors $p^\mu$ and $s^\mu$, and
going to the longitudinal polarization case $s^\mu \rightarrow
p^\mu/M$, one finds oneself in a more complicated situation. It
appears, that two pieces arise: the first one is related to the {\it
transverse} polarization ($g_{1+2}$ or
$c_T$) and the second to the {\it difference} between longitudinal and
transverse polarizations ($g_2$ or $b_A$).

This decomposition does not result in any physical effects, if only
the high-$Q^2$ region is considered. However when one goes to the
low-$Q^2$ region, there is a natural source for the $g_2$ strong
$Q^2$-dependence: namely, the elastic contribution to the BC sum rule.
This should not be mixed with the elastic contribution to the
dispersion integral itself \ci{Ji}. In this later case, the original
problem with the generalized GDH sum rule does not occur anymore.

Although there is no dynamical information about $g_{1+2}$ at low
$Q^2$, it is {\it possible} to suppose its smooth behaviour and it
leads to a zero for the GDH integral at $Q^2\sim 0.2 GeV^2$. This value
is just determined  by the elastic BC contribution and is not
affected by a possible BC violation induced by  Regge cuts.
The saturation of the GDH sum rule by  low-lying resonances leads
to substantially different values $Q^2 \sim 0.5-0.8 GeV^2$. It is then
natural, that the simultaneous use of the BC sum rule results in a
sharp and even oscillating behaviour of $I_{1+2}$. Moreover,
the $\Delta(1232)$ contribution to $g_1$ via $g_2$ supports
qualitatively a sharp $Q^2$-dependence of $I_2$. Further analysis
of the resonance contributions is strongly required and in particular,
it is very important to study systematically their decomposition into
$g_{1+2}$ and
$g_2$.

The flavour dependence of the GDH sum rule suggests, that its sharp
dependence is associated with the $SU(3)$ singlet channel. Therefore
a possible relation to the EMC Spin Crisis via the gluonic anomaly
seems to be of great interest.

\section*{Acknowledgments}

One of us (O.T.) is indebted to A.P.Bakulev,A.V.Efremov and R.Ruskov
for useful  discussions, to B.L.Ioffe for critical remarks and to
R.Workman for helpful correspondence.

\bb{99}
\bi{Fey} R.P.Feynman {\it Photon-Hadron Interactions}, (Benjamin,
Reading, MA, 1972).
\bi{Sch} J.Schwinger, Proc. Natl. Acad. Sci. U.S.A. {\bf 72},
(1975) 1559.
\bi{SoTe} J.Soffer and O.Teryaev, Phys. Rev. Lett. {\bf
70}, (1993) 3373.
\bi{Ji} Xiangdong Ji, Phys. Lett. {\bf B309 }, (1993) 187.
\bi{WoLi} R.Workman and Z.Li, Phys. Rev. Lett. {\bf 71}, (1993) 3608.
\bi{Ger} S.B.Gerasimov, Yad. Fiz. {\bf 2}, (1965) 598 [Sov. J. Nucl
Phys. {\bf 2}, (1966) 430].
\bi{DH} S.D.Drell and A.C.Hearn, Phys. Rev.
Lett. {\bf 16}, (1966) 908.
\bi{BC} H.Burkhardt and W.N.Cottingham,
Ann. Phys. (N.Y.) {\bf 56}, (1970) 453.
\bi{Rad} V.A.Nesterenko,
A.V.Radyushkin, Pis'ma ZhETF {\bf 39}, (1984) 576 [JETP Lett. {\bf 39},
(1984) 707].
\bi{SMC} B. Aveda et al., (Spin Muon Collaboration), Phys. Lett. {\bf
B302}, (1993) 533.
\bi{SLAC} P.L. Anthony et al., (E142 Collaboration), Phys. Rev.
Lett. {\bf 71}, (1993) 959.
\bi{Meis} V.Bernard, N.Kaiser and U.- G.Meissner, Phys. Rev.
{\bf D48}, (1993) 3062.
\bi{IoLi} B.L.Ioffe, V.A.Khoze, L.N.Lipatov, {\it Hard Processes},
 North-Holland, Amsterdam, 1984.
\bi{Jaf} R.L.Jaffe and Xiangdong Ji, Phys. Rev. {\bf D43}, (1991) 724.
\bi{BuLi} V. Burkert and Z. Li, Phys. Rev. {\bf D47}, (1993) 46; V.
 Burkert and B.L.Ioffe, Phys. Lett. {\bf B296}, (1992) 223.
\bi{ET84} A.Efremov and O.Teryaev, Yad. Fiz. {\bf 39}, (1984) 1517
[Sov. J. Nucl Phys. {\bf 39}, (1984) 962].
\bi{ET85} A.Efremov and O.Teryaev, Phys.
Lett. {\bf B150}, (1985) 383.
\bi{Ah} M.Ahmed and G.G.Ross, Nucl. Phys.
{\bf B111}, (1976) 441.
\bi{StQu} J.Qiu and G.Sterman, Nucl. Phys. {\bf
B378}, (1992) 52.
\bi{Mil} Wu-Yang Tsai, L.L. DeRaad, Jr. and K.A. Milton, Phys.
Rev. {\bf D11}, (1975) 3537.
\bi{EST} A.V.Efremov, J.Soffer, O.V.Teryaev, Nucl.Phys. {\bf B346},
(1990)  97.
\bi{Mer-vN} R. Mertig and W.L. van Neerven, Preprint INLO-PUB-2/93.
\bi{Pom} The Pomeron cut can also occur in the scaling region
as a correlator singularity at
$x \sim 0$, making the $g_2$ integral divergent, a  possibility
mentioned long ago in ref.\ci{Hei}.
\bi{Hei} R.L.Heimann, Nucl.Phys. {\bf B64}, (1973) 429.
\bi{JVer} J. Vermaseren, FORM user's Guide (Nikhef, Amsterdam, 1990).
\bi{OT92} O.V. Teryaev in {\it Proceedings of XI International Seminar
on High-Energy Physics Problems, Dubna, Sept.1992}.\eb
\end{document}